\documentclass[11pt]{article}
\usepackage{amssymb,amsmath,amsfonts}
\usepackage{graphicx}
\usepackage{graphics}
\usepackage{eepic,epsfig}
\usepackage{dcolumn}


\begin{document}

\title{Topological Casimir effect for fermionic condensate \\
in AdS spacetime with compact dimensions}
\author{T. A. Petrosyan, A. A. Grigoryan and A. A. Saharian \vspace{0.3cm}
\\
\textit{Institute of Physics, Yerevan State University,}\\
\textit{1 Alex Manoogian Street, 0025 Yerevan, Armenia} \vspace{0.3cm} }
\maketitle

\begin{abstract}
We investigate combined effects of gravitational field and spatial topology
on the fermionic condensate (FC) for a massive Dirac field in locally
anti-de Sitter (AdS) spacetime with a part of spatial dimensions
compactified to a torus. For general phases in the periodicity conditions
along compact dimensions the topological Casimir contribution is explicitly
extracted and the renormalization is reduced to the one for purely AdS
spacetime. The FC is an even periodic function of the magnetic flux enclosed
by compact dimension with the period of flux quantum. The topological
contribution vanishes on the AdS boundary and dominates in the total FC in
the region near the AdS horizon. For proper lengths of compact dimensions
smaller than the AdS curvature radius the influence of the gravitational
field is weak and the leading term in the corresponding expansion coincides
with the FC for a locally Minkowski spacetime with compact dimensions. For
large proper lengths the decay of the topological FC follows a power law for
both massless and massive field, in contrast to an exponential decay in
Minkowski bulk for massive fields. Applications are discussed for 2D Dirac
materials.
\end{abstract}

\bigskip


\bigskip

\section{Introduction}

The Casimir effect (for reviews see \cite{Most97,Milt02,Bord09,Dalv11}) is
sourced by conditions imposed on the operators of quantum fields in the
Heisenberg representation, in addition to the field equations. These
conditions can be categorized into two distinct classes. The first class
includes the conditions imposed on boundaries separating spacetime regions
with different physical properties. These boundaries may have different
physical natures. Examples that have been examined in the extant literature
include interfaces between two media, boundaries separating different phases
of a physical system, horizons in gravitational fields and in noninertial
reference frames, and branes in models with extra dimensions. The second
class of conditions is induced by nontrivial spatial topology. The
periodicity conditions along compact dimensions modify the spectrum of
fluctuations of quantum fields giving rise to shifts in the expectation
values of physical observables depending on the local geometry and
compactification scheme. This is the essence of the topological Casimir
effect.

The topological Casimir effect has been a subject of considerable interest
as a possible mechanism for the stabilization of moduli fields in
Kaluza-Klein-type theories with extra spatial dimensions and in string
theories. The topological contribution to the vacuum energy density has also
been examined as a possible origin of dark energy responsible for the
accelerating expansion of the universe at its current stage \cite%
{Milt03,Eliz06,Dupa13,Wong15}. In the present paper we discuss the effects
of nontrivial topology on the vacuum fermion condensate (FC) $\left\langle 
\bar{\psi}\psi \right\rangle $ for a Dirac field $\psi $ in a locally
anti-de Sitter (AdS) spacetime with toroidally compactified spatial
dimensions. In addition to the energy-momentum tensor and current density
for charged fields, the FC is an important local characteristic of the
vacuum state, bilinear in the field operator. The Fermi condensation is
among the most interesting phenomena in condensed matter physics. Its
applications include important areas such as superconductivity and phase
transitions, dynamical mass generation and symmetry breaking, quantum optics
and quantum chromodynamics. A plethora of mechanisms for the formation of
the FC have been contemplated in the extant literature. These include
diverse forms of interactions among fermionic particles and quasiparticles,
particularly the Nambu-Jona-Lasinio (NJL)-type models with self-interacting
fields. An intriguing avenue for exploration in the study of Fermi
condensation pertains to the dependence of the FC on the characteristics of
background geometry (see, e.g., \cite{Inag97}-\cite{Saha21}). In particular,
the boundary and periodicity conditions responsible for the Casimir effect,
have the capacity to either reduce or enlarge the parameter space for phase
transitions.

In our consideration, the AdS spacetime has been selected as the background
geometry due to its significance in a number of recent advancements in
gravitational physics and quantum field theory. The maximal symmetry of the
AdS geometry allows to obtain exact results in a large number of physical
problems and the problem under consideration in the present paper is one of
those examples. The properties of the AdS spacetime play a crucial role in
the formulations of braneworld models with large extra dimensions \cite%
{Maar10} and in AdS/CFT correspondence \cite{Ammo15}. The latter establishes
a holographic duality between supergravity and string theories on AdS bulk
from one side and conformal field theory living on its boundary from
another. In addition to high energy physics, this duality has interesting
applications in condensed matter physics. Motivated by the need for
stabilization of radion field and generation of cosmological constant on
branes, the Casimir effect in braneworlds on AdS bulk has been widely
studied in the literature (see references in \cite{SahaRev20}). The
renormalized FC in AdS spacetime has been considered in \cite{Ambr15,Ambr17}
at zero and finite temperatures. The contributions induced by boundaries
(branes) and cosmic string-type defects were discussed in \cite{Beze13}-\cite%
{Bell22}.

The organization of the present paper is as follows. In the next section the
problem is formulated and the complete set of fermionic modes is presented.
By using these modes, in the third section we evaluate the renormalized FC.
The asymptotics near the AdS boundary and near the horizon are investigated
and limiting cases are discussed in the fourth section. The FC in $C$-, $P$%
-, and $T$-symmetric odd-dimensional models is considered. Applications are
discussed in the structures of 2D Dirac materials with nontrivial topology.
The main results are summarized in the last section.

\section{Problem setup and the fermionic modes}

\label{sec:Setup}

We consider $(D+1)$-dimensional locally AdS spacetime with the metric tensor 
$g_{\mu \nu }$ in coordinates $x^{\mu }=(x^{0},\mathbf{x},x^{D})$, where $%
\mathbf{x}=(x^{1},x^{2},\cdots ,x^{D-1})$. The geometry is sourced by a
negative cosmological constant $\Lambda $ and the corresponding Riemann
tensor reads $R_{\ \nu \alpha \beta }^{\mu }=(\delta _{\beta }^{\mu }g_{\nu
\alpha }-\delta _{\alpha }^{\mu }g_{\nu \beta })/a^{2}$ with the curvature
radius given by $a=\sqrt{D(1-D)/(2\Lambda )}$. In the Poincar\'{e}
coordinates with $x^{D}=z$, the line element is written in conformally flat
form%
\begin{equation}
ds^{2}=g_{\mu \nu }dx^{\mu }dx^{\nu }=\frac{a^{2}}{z^{2}}\left( dt^{2}-d%
\mathbf{x}^{2}-dz^{2}\right) ,  \label{ds2}
\end{equation}%
where $d\mathbf{x}^{2}=\sum_{n=1}^{D-1}(dx^{n})^{2}$. The coordinate $z$
varies in the range $z\in \lbrack 0,\infty )$ and the hypersurfaces $z=0$
and $z=\infty $ correspond to the AdS boundary and horizon, respectively.
The global geometry we are interested in differs from that for AdS
spacetime. It will be assumed that the subspace covered by the coordinates $%
\mathbf{x}_{q}=\left( x^{p+1},\cdots ,x^{D-1}\right) $ is compactified on
the torus $T^{q}=(S^{1})^{q}$, $q=D-p-1$, with the lengths of compact
dimensions $(L_{p+1},\cdots ,L_{D-1})$: $x^{l}\in \lbrack 0,L_{l}]$, $%
l=p+1,\cdots ,D-1$. For the remaining spatial coordinates $\mathbf{x}%
_{p}=\left( x^{1},\cdots ,x^{p}\right) $ one has, as usual, $x^{l}\in
(-\infty ,+\infty )$, $l=1,\cdots ,p$. The physical distance between the
points along the $z$-axis is measured by the new coordinate $y$ defined as $%
y=a\ln (z/a)$ with the variation range $y\in (-\infty ,+\infty )$. For the
proper lengths of the compact dimensions, $L_{\mathrm{(p)}l}$, measured by
an observer with fixed coordinates $(\mathbf{x},x^{D})$, one has $L_{\mathrm{%
(p)}l}=aL_{l}/z=e^{-y/a}L_{l}$.

Having described the background geometry we turn to the field content. We
will consider a massive Dirac field $\psi \left( x\right) $ in the presence
of an external classical Abelian gauge field $A_{\mu }\left( x\right) $
(here $x$ is used for the collective notation of the spacetime coordinates $%
x^{\mu }$). The dynamics of the field in the Heisenberg picture is described
by the equation 
\begin{equation}
\left( i\gamma ^{\mu }D_{\mu }-sm\right) \psi (x)=0,\;D_{\mu }=\partial
_{\mu }+\Gamma _{\mu }+ieA_{\mu },  \label{deq}
\end{equation}%
with $m$ and $e$ being the mass and the charge of the field quanta. The spin
connection $\Gamma _{\mu }$ is expressed in terms of the curved spacetime
Dirac matrices $\gamma ^{\mu }$ as $\Gamma _{\mu }=\frac{1}{4}\gamma ^{\nu
}\nabla _{\mu }\gamma _{\nu }$, where $\gamma _{\nu }=g_{\nu \alpha }\gamma
^{\alpha }$ and $\nabla _{\mu }$ stands for the standard covariant
derivative acting on spacetime vectors $\gamma _{\nu }$. In the irreducible
representation, the Clifford algebra is realized by the $N\times N$ matrices 
$\gamma ^{\mu }$, where $N=2^{\left[ \left( D+1\right) /2\right] }$ and $%
\left[ x\right] $ is the integer part of $x$ (for the Dirac matrices in an
arbitrary number of the spacetime dimension, see \cite{Park09}). In
even-dimensional spacetimes (odd values of $D$) the parameter $s$ in (\ref%
{deq}) is taken $s=+1$. In odd number of spacetime dimensions (even values
of $D$) $s$ takes the values $s=+1$ and $s=-1$ and they correspond to two
inequivalent irreducible representations of the Clifford algebra (see below).

Here we use the representation of the Dirac matrices given by 
\begin{equation}
\gamma ^{0}=\frac{z}{a}\left( 
\begin{array}{cc}
0 & \chi _{0} \\ 
\chi _{0}^{\dagger } & 0%
\end{array}%
\right) ,\,\boldsymbol{\gamma }=\frac{z}{a}\left( 
\begin{array}{cc}
0 & \boldsymbol{\chi } \\ 
-\boldsymbol{\chi }^{\dagger } & 0%
\end{array}%
\right) ,  \label{dmf}
\end{equation}%
with $\boldsymbol{\gamma }=(\gamma ^{1},\cdots \gamma ^{D-1})$, $\boldsymbol{%
\chi }=(\chi _{1},\cdots ,\chi _{D-1})$, and $\,\gamma ^{D}=i(z/a)\mathrm{%
diag}(1,-1)$. Here, $\chi _{l}$, $l=0,1,\cdots ,D-1$, are constant $%
N/2\times N/2$ matrices obeying the relations $\chi _{0}\chi _{l}^{\dagger
}=\chi _{l}\chi _{0}^{\dagger }$, $\chi _{0}^{\dagger }\chi _{l}=\chi
_{l}^{\dagger }\chi _{0}$, $\chi _{0}^{\dagger }\chi _{0}=1$, and $\chi
_{l}\chi _{n}^{\dagger }+\chi _{n}\chi _{l}^{\dagger }=\chi _{l}^{\dagger
}\chi _{n}+\chi _{n}^{\dagger }\chi _{l}=2\delta _{nl}$ for $l,n=1,2,\ldots
,D-1$. These properties follow from the anticommutation relation $\{\gamma
^{\mu },\gamma ^{\nu }\}=2g^{\mu \nu }$. With this choice, for the
corresponding components of the spin connection one gets%
\begin{equation}
\Gamma _{\mu }=\frac{\gamma ^{D}}{2z}\gamma _{\mu },\;\Gamma _{D}=0,\,\mu
=0,1,\cdots ,D-1.  \label{Gamu}
\end{equation}%
In the special case $D=2$ we have $\chi _{0}=\chi _{1}=1$ and $\gamma
^{0}=z\sigma _{\text{\textrm{P}}1}/a$, $\gamma ^{1}=iz\sigma _{\text{\textrm{%
P}}2}/a$, $\gamma ^{2}=iz\sigma _{\text{\textrm{P}}3}/a$, where $\sigma _{%
\text{\textrm{P}}\mu }$ are the Pauli matrices.

The spatial topology under consideration is nontrivial and to complete the
problem formulation, it is also necessary to specify the periodicity
conditions along compact directions. We will impose quasiperiodic conditions
with general constant phases $\alpha _{l}$, given by 
\begin{equation}
\psi \left( t,\mathbf{x}_{p},\mathbf{x}_{q}+L_{l}\mathbf{e}_{(l)},z\right)
=e^{i\alpha _{l}}\psi \left( t,\mathbf{x}_{p},\mathbf{x}_{q},z\right) ,
\label{qpc}
\end{equation}%
where $\mathbf{e}_{(l)}$, $l=p+1,\cdots ,D-1$, is the unit vector in the
subspace $\mathbf{x}_{q}$ directed along the $l$-th compact coordinate $x^{l}
$ with the components $e_{(l)}^{n}=\delta _{l}^{n}$. The untwisted and
twisted fields are special cases with the phases $\alpha _{l}=0$ and $\alpha
_{l}=\pi $, respectively. Instead of the fields $\{\psi (x),A_{\mu }(x)\}$
we can consider a new set $\{\psi ^{\prime }(x),A_{\mu }^{\prime }(x)\}$
obtained by the gauge transformation 
\begin{equation}
\psi ^{\prime }(x)=\psi (x)e^{ieg(x)},\,A_{\mu }^{\prime }(x)=A_{\mu
}(x)-\partial _{\mu }g(x),  \label{gaugeTr}
\end{equation}%
with some function $g(x)$. The field equation is form-invariant under this
transformation and the periodicity condition for the new field reads%
\begin{equation}
\psi ^{\prime }\left( t,\mathbf{x}_{p},\mathbf{x}_{q}+L_{l}\mathbf{e}%
_{(l)},z\right) =\psi ^{\prime }\left( t,\mathbf{x}_{p},\mathbf{x}%
_{q},z\right) e^{i\alpha _{l}+ie\left[ g(t,\mathbf{x}_{p},\mathbf{x}%
_{q}+L_{l}\mathbf{e}_{(l)},z)-g(t,\mathbf{x}_{p},\mathbf{x}_{q},z)\right] }.
\label{qpc2}
\end{equation}%
In what follows we will consider a gauge field configuration with constant
covariant components $A_{\mu }=\mathrm{const}$. This field is removed from
the field equation passing to a new gauge (\ref{gaugeTr}) with the function $%
g(x)=A_{\mu }x^{\mu }$ and $A_{\mu }^{\prime }(x)=0$. With this choice, the
periodicity condition for the field $\psi ^{\prime }(x)$ takes the form 
\begin{equation}
\psi ^{\prime }\left( t,\mathbf{x}_{p},\mathbf{x}_{q}+L_{l}\mathbf{e}%
_{(l)},z\right) =e^{i\tilde{\alpha}_{l}}\psi ^{\prime }\left( t,\mathbf{x}%
_{p},\mathbf{x}_{q},z\right) ,  \label{qpc3}
\end{equation}%
where the new phases are defined as 
\begin{equation}
\tilde{\alpha}_{l}=\alpha _{l}+eA_{l}L_{l}.  \label{alphal}
\end{equation}%
The FC will depend on the phases $\alpha _{l}$ and on the constant
components of the vector potential along compact dimensions through the
combination $\tilde{\alpha}_{l}$. The physical relevance of the constant
gauge field in the problem under consideration is related to the nontrivial
spatial topology which leads to an Aharonov-Bohm-type effect. The phase
shift in (\ref{alphal}), induced by the gauge field, is presented as $%
eA_{l}L_{l}=-2\pi \Phi _{l}/\Phi _{0}$, where $\Phi _{0}=2\pi /e$ is the
flux quantum. The quantity $\Phi _{l}$ can be formally interpreted as the
magnetic flux enclosed by the $l$-th compact dimension. In the discussion
below we will work in the gauge $\{\psi ^{\prime }(x),A_{\mu }^{\prime }=0\}$
with the periodicity condition (\ref{qpc3}) omitting the prime.

For the evaluation of the FC we need the complete set of positive and
negative energy fermionic modes $\psi _{\beta }^{\left( \pm \right) }(x)$
obeying the field equation (\ref{deq}) with $A_{\mu }=0$ and the condition (%
\ref{qpc3}). These modes have been considered in \cite{Bell17} for two
different representations of the Dirac matrices. The modes are specified by
the set of quantum numbers $\beta =(\mathbf{k},\lambda ,\sigma )$ where $%
0\leq \lambda <\infty $ and $\sigma =1,2,\cdots ,N/2$. The vector $\mathbf{k}%
=(k_{1},k_{2},\cdots ,k_{D-1})$ is the momentum in the subspace $%
(x^{1},x^{2},\cdots ,x^{D-1})$. It is decomposed into two parts: the part $%
\mathbf{k}_{(p)}=(k_{1},k_{2},\cdots ,k_{p})$ with the components $-\infty
<k_{l}<+\infty $ for $l=1,\cdots ,p$, and the part $\mathbf{k}%
_{(q)}=(k_{p+1},\cdots ,k_{D-1})$. The components of the latter are
quantized by the condition (\ref{qpc3}) and take the values 
\begin{equation}
k_{l}=\frac{2\pi n_{l}+\tilde{\alpha}_{l}}{L_{l}},\,l=p+1,\ldots ,D-1,
\label{kl}
\end{equation}%
with $n_{l}=0,\pm 1,\pm 2,\cdots $. The positive and negative energy modes
are expressed as%
\begin{equation}
\psi _{\beta }^{\left( \pm \right) }(x)=\frac{a^{-\frac{D}{2}}z^{\frac{D+1}{2%
}}}{2\left( 2\pi \right) ^{\frac{p}{2}}}\sqrt{\frac{\lambda }{V_{q}}}e^{i%
\mathbf{kx}\mp i\omega t}B_{\beta }^{(\pm )}(z),  \label{psi}
\end{equation}%
where $V_{q}=L_{p+1}\cdots L_{D-1}$ is the coordinate volume of the compact
subspace, $\omega =\sqrt{\lambda ^{2}+k^{2}}$, $\mathbf{kx}%
=\sum_{l=1}^{D-1}k_{l}x^{l}$, and the $N$-component spinors $B_{\beta
}^{(\pm )}(z)$ are given by%
\begin{align}
B_{\beta }^{(+)}(z) &=\left( 
\begin{array}{c}
\frac{\mathbf{k}\boldsymbol{\chi } \chi _{0}^{\dag }+i\lambda -\omega }{%
\omega }w^{\left( \sigma \right) }J_{ma+\frac{s}{2}}\left( \lambda z\right)
\\ 
si\chi _{0}^{\dag }\frac{\mathbf{k} \boldsymbol{\chi } \chi _{0}^{\dag
}+i\lambda +\omega }{\omega }w^{\left( \sigma \right) }J_{ma-\frac{s}{2}%
}\left( \lambda z\right)%
\end{array}%
\right) ,  \notag \\
B_{\beta }^{(-)}(z) &=\left( 
\begin{array}{c}
i\chi _{0}\frac{\mathbf{k} \boldsymbol{\chi }^{\dag }\chi _{0}-i\lambda
+\omega }{\omega }w^{\left( \sigma \right) }J_{ma+\frac{s}{2}}\left( \lambda
z\right) \\ 
s\frac{\mathbf{k} \boldsymbol{\chi }^{\dag }\chi _{0}-i\lambda -\omega }{%
\omega }w^{\left( \sigma \right) }J_{ma-\frac{s}{2}}\left( \lambda z\right)%
\end{array}%
\right) .  \label{Be}
\end{align}%
In these expressions $\mathbf{k} \boldsymbol{\chi }=\sum_{l=1}^{D-1}k_{l}%
\chi _{l}$, $J_{\nu }(u)$ is the Bessel function, and $w^{\left( \sigma
\right) }$ are one-column matrices having $N/2$ rows with the elements $%
w_{l}^{\left( \sigma \right) }=\delta _{l\sigma }$. For the modes given in 
\cite{Bell17} the factor $s$ in the second lines of (\ref{Be}) is absent.
This is related to the different choice of the gamma matrix $\gamma ^{D}$
for $s=-1$.

\section{Fermionic condensate: General formula}

\label{sec:FC}

The ground state FC is defined as the vacuum expectation value $\langle 0|%
\bar{\psi}\psi |0\rangle \equiv \langle \bar{\psi}\psi \rangle $, where $%
|0\rangle $ stands for the vacuum state and $\bar{\psi}(x)=\psi ^{\dag
}(x)\gamma ^{(0)}$ is the Dirac adjoint with $\gamma ^{(0)}=a\gamma ^{0}/z$
being the flat spacetime matrix. Expanding the fermionic operator in terms
of the modes $\psi _{\beta }^{\left( \pm \right) }(x)$ and using the
anticommutation relations for the annihilation and creation operators, the
following mode sum is obtained: 
\begin{equation}
\langle \bar{\psi}\psi \rangle =-\frac{1}{2}\sum_{\beta }\sum_{\kappa =\pm
}\kappa \bar{\psi}_{\beta }^{\left( \kappa \right) }\left( x\right) \psi
_{\beta }^{\left( \kappa \right) }\left( x\right) .  \label{FCms}
\end{equation}%
Here, $\sum_{\beta }$ is understood as the summation for discrete subset of
quantum numbers ($\mathbf{k}_{(q)}$, $\sigma $) and as an integration over
the continuous ones ($\mathbf{k}_{(p)}$, $\lambda $). The expression on the
right-hand side in (\ref{FCms}) is divergent. It is assumed that some
regularization procedure has been applied without explicitly writing it. We
are interested in the topological contribution which is finite and the
procedure for its extraction from the total FC, described below, does not
depend on the way of regularization.

With the modes (\ref{psi}), we can evaluate the FC by using the mode sum
formula (\ref{FCms}). By taking into account that for a $N/2\times N/2$
matrix $M$ we have $\sum_{\sigma }w^{\left( \sigma \right) \dag }Mw^{\left(
\sigma \right) }=\mathrm{Tr}(M)$, the FC is presented as 
\begin{equation}
\langle \bar{\psi}\psi \rangle =-\frac{sNz^{D+1}}{2V_{q}a^{D}}\underset{%
\mathbf{n}_{q}}{\sum }\int \frac{d\mathbf{k}_{\left( p\right) }}{\left( 2\pi
\right) ^{p}}\int_{0}^{\infty }d\lambda \frac{\lambda ^{2}}{\omega }J_{ma-%
\frac{1}{2}}\left( \lambda z\right) J_{ma+\frac{1}{2}}\left( \lambda
z\right) ,  \label{FC1}
\end{equation}%
where $\mathbf{n}_{q}=\left( n_{p+1},\ldots ,n_{D-1}\right) $ and $\sum_{%
\mathbf{n}_{q}}=\sum_{n_{p+1}=-\infty }^{+\infty }\cdots \sum_{n_{D}=-\infty
}^{+\infty }$. As seen from (\ref{FC1}), the FC has opposite signs for the
cases $s=+1$ and $s=-1$. The further consideration will be presented for the
case $s=+1$. In order to transform the multiple integral in the right-hand
side of (\ref{FC1}), firstly, we will pass from the integration over $%
\mathbf{k}_{\left( p\right) }$ to an integration over the hyperspherical
coordinates in the momentum space. By subsequent integration over the
angular part the expression (\ref{FC1}) is reduced to 
\begin{equation}
\langle \bar{\psi}\psi \rangle =-\frac{Na^{-D}z^{D+1}}{\left( 4\pi \right) ^{%
\frac{p}{2}}\Gamma \left( \frac{p}{2}\right) V_{q}}\underset{\mathbf{n}_{q}}{%
\sum }\int_{0}^{\infty }dk_{\left( p\right) }k_{\left( p\right)
}^{p-1}\int_{0}^{\infty }d\lambda \frac{\lambda ^{2}}{\omega }J_{ma-\frac{1}{%
2}}\left( \lambda z\right) J_{ma+\frac{1}{2}}\left( \lambda z\right) .
\label{FC2}
\end{equation}%
The corresponding expression of the FC for a massive fermionic field on the
background of AdS spacetime with trivial spatial topology, in the presence
of two parallel boundaries with bag boundary conditons on them, is given in 
\cite{Eliz13}.

By using the representation $\sqrt{\pi }/\omega =\int_{0}^{\infty
}dv\,e^{-\omega ^{2}v}/\sqrt{v}$, the integration over $\lambda $ in (\ref%
{FC2}) is done by using the formula 
\begin{equation}
\int_{0}^{\infty }d\lambda \lambda ^{2}e^{-\lambda ^{2}v}J_{\nu -\frac{1}{2}%
}\left( \lambda z\right) J_{\nu +\frac{1}{2}}\left( \lambda z\right) =\frac{%
ze^{-u}}{4v^{2}}\left[ I_{\nu -\frac{1}{2}}\left( u\right) -I_{\nu +\frac{1}{%
2}}\left( u\right) \right] ,  \label{Int1}
\end{equation}%
where $u=z^{2}/2v$ and $I_{\mu }\left( u\right) $ is the modified Bessel
function. This formula is obtained from the integral with the function $%
J_{\nu +1/2}^{2}\left( \lambda z\right) $ instead of the product of the
Bessel functions, given in \cite{Prud2}, by employing the recurrence
relation for the Bessel functions. After evaluating the integral over $%
k_{\left( p\right) }$ we get 
\begin{equation}
\langle \bar{\psi}\psi \rangle =\frac{Na^{-D}z^{q}}{2\left( 2\pi \right) ^{%
\frac{p+1}{2}}V_{q}}\int_{0}^{\infty }du\,u^{\frac{p+1}{2}}e^{-u}\sum_{j=\pm
1}jI_{ma+\frac{j}{2}}\left( u\right) \underset{\mathbf{n}_{q}}{\sum }e^{-%
\frac{k_{\left( q\right) }^{2}z^{2}}{2u}},  \label{FC3}
\end{equation}%
with $k_{\left( q\right) }^{2}=\sum_{i=p+1}^{D-1}k_{i}^{2}$. An alternative
representation of the FC is obtained by using the formula 
\begin{equation}
\overset{+\infty }{\underset{n_{l}=-\infty }{\sum }}e^{-vk_{l}^{2}}=\frac{%
L_{l}}{2\sqrt{\pi v}}\overset{+\infty }{\underset{n_{l}=-\infty }{\sum }}%
e^{in_{l}\tilde{\alpha}_{l}-\frac{L_{l}^{2}n_{l}^{2}}{4v}},  \label{Presum}
\end{equation}%
which is obtained applying the Poisson resummation formula to the left-hand
side. Introducing the notation $g_{\mathbf{n}_{q}}^{2}=%
\sum_{j=p+1}^{D-1}n_{j}^{2}L_{j}^{2}$, the FC is presented in the form 
\begin{equation}
\langle \bar{\psi}\psi \rangle =-\frac{Na^{-D}}{2\left( 2\pi \right) ^{\frac{%
D+1}{2}}}\underset{\mathbf{n}_{q}}{\sum }e^{i\tilde{\boldsymbol{\alpha }}%
\cdot \mathbf{n}_{q}}U_{ma}^{\frac{D+1}{2}}\left( b_{\mathbf{n}_{q}}\right) ,
\label{FC4}
\end{equation}%
with $\tilde{\boldsymbol{\alpha }}\cdot \mathbf{n}_{q}\equiv
\sum_{j=p+1}^{D-1}n_{j}\tilde{\alpha}_{j}$ and $b_{\mathbf{n}_{q}}=1+g_{%
\mathbf{n}_{q}}^{2}/(2z^{2})$. In (\ref{FC4}) we have defined the function 
\begin{equation}
U_{\nu }^{\mu }\left( x\right) =\sqrt{2\pi }\int_{0}^{\infty }du\,\frac{%
u^{\mu -\frac{1}{2}}}{e^{xu}}\left[ I_{\nu -\frac{1}{2}}\left( u\right)
-I_{\nu +\frac{1}{2}}\left( u\right) \right] .  \label{px}
\end{equation}%
For $\nu >-1/2$ this function is positive and monotonically decreases with
increasing $x$ in the region $x>1$. The integral in (\ref{px}) is expressed
in terms of the associated Legendre function of the second kind $Q_{\nu
}^{\mu }\left( u\right) $ (see \cite{Prud2}) and we obtain an equivalent
representation 
\begin{equation}
U_{\nu }^{\mu }\left( x\right) =2e^{-\mu \pi i}\frac{Q_{\nu -1}^{\mu }\left(
x\right) -Q_{\nu }^{\mu }\left( x\right) }{\left( x^{2}-1\right) ^{\mu /2}}.
\label{pQ}
\end{equation}

A simple expression for the function (\ref{pQ}) is obtained for $\mu =n+1/2$
with $n=1,2,\ldots $ (corresponding to spatial dimension $D=2n$ in (\ref{FC4}%
)): 
\begin{equation}
U_{\nu }^{n+\frac{1}{2}}\left( x\right) =\left( -1\right) ^{n}2^{\nu +1}%
\sqrt{\pi }\partial _{x}^{n}\left[ \frac{1/\sqrt{x+1}}{\left( \sqrt{x+1}+%
\sqrt{x-1}\right) ^{2\nu }}\right] .  \label{pnuu}
\end{equation}%
For integer values of $\mu $, $\mu =n$ (odd number of spatial dimensions in (%
\ref{FC4}), $D=2n-1$), by using the formula from \cite{Nist10} for the
associated Legendre functions in (\ref{pQ}), we get 
\begin{equation}
U_{\nu }^{n}\left( x\right) =2\left( -1\right) ^{n}\partial _{x}^{n}\left[
Q_{\nu -1}\left( x\right) -Q_{\nu }\left( x\right) \right] ,  \label{pu2}
\end{equation}%
with $Q_{\nu }\left( x\right) $ being the Legendre function of the second
kind.

The divergent part of the FC is contained in the term of (\ref{FC4}) with $%
n_{l}=0$ for all $l=p+1,\cdots ,D-1$. This term corresponds to the
condensate in non-compactified AdS spacetime. It is given by the expression 
\begin{equation}
\langle \bar{\psi}\psi \rangle _{0}=\frac{-Na^{-D}}{2\left( 2\pi \right) ^{%
\frac{D}{2}}}\int_{0}^{\infty }du\,\frac{u^{\frac{D}{2}}}{e^{u}}\left[ I_{ma-%
\frac{1}{2}}\left( u\right) -I_{ma+\frac{1}{2}}\left( u\right) \right] .
\label{FC0}
\end{equation}%
For $D=4$ this expression coincides with the result found in \cite{Bell22}.
The expression in the right hand side of (\ref{FC0}) needs regularization
with further renormalization. The renormalization of the FC in AdS spacetime
with $D=4$, described in global coordinates, has been discussed in \cite%
{Ambr15}. Here we are interested in the topological (Casimir) contribution $%
\langle \bar{\psi}\psi \rangle _{\text{\textrm{c}}}=\langle \bar{\psi}\psi
\rangle -\langle \bar{\psi}\psi \rangle _{0}$ given by 
\begin{equation}
\langle \bar{\psi}\psi \rangle _{\text{\textrm{c}}}=-\frac{Na^{-D}}{2\left(
2\pi \right) ^{\frac{D+1}{2}}}\sideset{}{'}{\sum}_{\mathbf{n}_{q}}\cos
\left( \tilde{\boldsymbol{\alpha }}\cdot \mathbf{n}_{q}\right) U_{ma}^{\frac{%
D+1}{2}}\left( b_{\mathbf{n}_{q}}\right) ,  \label{FCc}
\end{equation}%
where the prime on the summation sign means that the term with $n_{i}=0$, $%
i=p+1,\ldots ,D-1$, should be excluded from the sum. The topological part is
finite and the renormalization is required only for the pure AdS part $%
\langle \bar{\psi}\psi \rangle _{0}$. The topological Casimir contribution (%
\ref{FCc}) is an even periodic function of the magnetic flux enclosed by
compact dimensions. \ It depends on the lengths of compact dimensions $L_{l}$
and on the $z$-coordinate in the form of the ratio $L_{l}/z=L_{\mathrm{(p)}%
l}/a$ which presents the proper length of the compact dimension measured in
units of the curvature radius.

\section{Special cases, asymptotics, and numerical analysis}

\label{sec:Numer}

We start the consideration of special cases from the flat spacetime limit $%
a\rightarrow \infty $ for fixed $y$. In order to find the asymptotic
expression for the function $U_{ma}^{\frac{D+1}{2}}\left( b_{\mathbf{n}%
_{q}}\right) $ in (\ref{FC4}) we note that in the limit under consideration $%
ma\gg 1$ and $b_{\mathbf{n}_{q}}-1\ll 1$. By using the uniform asymptotic
expansion of the functions $I_{\nu \pm 1/2}\left( u\right) $ for large
values of the order, the integral in (\ref{px}) is expressed in terms of the
modified Bessel function $K_{\frac{D-1}{2}}\left( mg_{\mathbf{n}_{q}}\right) 
$ and from (\ref{FC4}) we obtain the FC\ in locally Minkowski spacetime with
spatial topology $R^{p+1}\times T^{q}$: 
\begin{equation}
\langle \bar{\psi}\psi \rangle _{\text{\textrm{c}}}^{\text{\textrm{M}}}=-%
\frac{Nm^{\frac{D+1}{2}}}{\left( 2\pi \right) ^{\frac{D+1}{2}}}%
\sideset{}{'}{\sum}_{\mathbf{n}_{q}}\cos \left( \tilde{\boldsymbol{\alpha }}%
\cdot \mathbf{n}_{q}\right) \frac{K_{\frac{D-1}{2}}\left( mg_{\mathbf{n}%
_{q}}\right) }{g_{\mathbf{n}_{q}}^{\frac{D-1}{2}}}.  \label{FCM}
\end{equation}%
This formula agrees with the result obtained in \cite{Bell09}. For a
massless field the condensate$\langle \bar{\psi}\psi \rangle _{\text{\textrm{%
c}}}^{\text{\textrm{M}}}$ vanishes.

The general expression (\ref{FCc}) is further simplified in the case of a
massless field. By taking into account that $U_{0}^{\mu }\left( u\right)
=2\Gamma \left( \mu \right) /\left( u+1\right) ^{\mu }$, for $m=0$ from (\ref%
{FCc}) one gets 
\begin{equation}
\langle \bar{\psi}\psi \rangle _{\text{\textrm{c}}}=-\frac{N\Gamma \left( 
\frac{D+1}{2}\right) }{\pi ^{\frac{D+1}{2}}a^{D}}\sideset{}{'}{\sum}_{%
\mathbf{n}_{q}}\frac{\cos \left( \tilde{\boldsymbol{\alpha }}\cdot \mathbf{n}%
_{q}\right) }{\left( 4+g_{\mathbf{n}_{q}}^{2}/z^{2}\right) ^{\frac{D+1}{2}}}.
\label{FCm0}
\end{equation}%
Unlike the case of locally Minkowski spacetime, the FC for a massless field
in AdS bulk is nonzero. This does not contradict the fact that in the
massless case the Dirac field is conformally invariant and the problem under
consideration is conformally related to the problem in Minkowski bulk
described by the line element $ds^{2}=dt^{2}-d\mathbf{x}^{2}-dz^{2}$. The
important point to note is that we have a conformal relation with the
problem formulated in the part $z\geq 0$ of Minkowski bulk and the
Minkowskian counterpart of the Dirac field obeys the bag boundary condition
at $z=0$. It is seen from the structure of the modes (\ref{psi}). For a
massless field we have $\psi _{\beta }^{\left( \pm \right)
}(x)=(z/a)^{D/2}\psi _{\mathrm{M},\beta }^{\left( \pm \right) }(x)$ with $%
\psi _{\mathrm{M},\beta }^{\left( \pm \right) }(x)$ being the corresponding
Minkowskian modes. By using the expressions (\ref{Be}), we can see that
those modes obey the bag boundary condition at $z=0$. Hence, the expected
conformal relation $\langle \bar{\psi}\psi \rangle _{\text{\textrm{c}}%
}=(z/a)^{D}\langle \bar{\psi}\psi \rangle _{\text{\textrm{c(1)}}}^{\mathrm{M}%
}$ takes place, where $\langle \bar{\psi}\psi \rangle _{\text{\textrm{c(1)}}%
}^{\mathrm{M}}$ is the FC in a locally Minkowski spacetime with a boundary
at $z=0$ on which the field obeys the bag boundary condition. The expression
for the latter is obtained from the right-hand side of (\ref{FCm0}).

As it has been mentioned above, the topological part in the FC depends on $%
L_{l}$ in the form of the ratio $L_{l}/z$. Let us consider the asymptotics
for small and large values of that ratio. For $L_{l}/z\ll 1$, corresponding
to small proper lengths of compact dimensions compared to the curvature
radius, we use the asymptotic $U_{\nu }^{\mu }\left( x\right) \approx \nu
\Gamma \left( \mu -1\right) \left( x-1\right) ^{1-\mu }$ for $x-1\ll 1$.
This formula is obtained with the help of (\ref{pQ}), taking into account
the next-to-leading term in the corresponding asymptotic expansion of the
Legendre function of the second kind. Thus, to the leading order we find $%
\langle \bar{\psi}\psi \rangle _{\text{\textrm{c}}}\approx
(z/a)^{D-1}\langle \bar{\psi}\psi \rangle _{\text{\textrm{c}}}^{\text{%
\textrm{M}}}$ with 
\begin{equation}
\langle \bar{\psi}\psi \rangle _{\text{\textrm{c}}}^{\text{\textrm{M}}%
}\approx -\frac{Nm\Gamma \left( \frac{D-1}{2}\right) }{4\pi ^{\frac{D+1}{2}}}%
\sideset{}{'}{\sum}_{\mathbf{n}_{q}}\frac{\cos \left( \tilde{\boldsymbol{%
\alpha }}\cdot \mathbf{n}_{q}\right) }{g_{\mathbf{n}_{q}}^{D-1}}.
\label{FCMm}
\end{equation}%
The expression in the right-hand side of (\ref{FCMm}) is derived from (\ref%
{FCM}) for $mL_{l}\ll 1$. We see that for $L_{l}/z\ll 1$ the leading term in
the expansion of $\langle \bar{\psi}\psi \rangle _{\text{\textrm{c}}}$ is
obtained from the corresponding result for the Minkowski bulk by the
replacement $L_{l}\rightarrow L_{\mathrm{(p)}l}$. Near the horizon or for
small lengths of compact dimensions the dominant contribution to the FC
comes from the fluctuations with small wavelengths and the effects induced
by the spacetime curvature are small.

The opposite limit $L_{l}/z\gg 1$ corresponds to large proper lengths of
compact dimensions or to the points near the AdS boundary. In this limit the
argument of the function $U_{ma}^{(D+1)/2}\left( b_{\mathbf{n}_{q}}\right) $
in (\ref{FCc}) is large. By using the corresponding asymptotic expression we
get 
\begin{equation}
\langle \bar{\psi}\psi \rangle _{\text{\textrm{c}}}\approx -\frac{N\Gamma
\left( ma+\frac{D+1}{2}\right) }{\pi ^{\frac{D}{2}}a^{D}\Gamma \left( ma+%
\frac{1}{2}\right) }\sideset{}{'}{\sum}_{\mathbf{n}_{q}}\frac{\cos \left( 
\tilde{\boldsymbol{\alpha }}\cdot \mathbf{n}_{q}\right) }{(g_{\mathbf{n}%
_{q}}/z)^{D+1+2ma}}.  \label{FCnb1}
\end{equation}%
As seen, the topological FC vanishes on the AdS boundary as $z^{D+1+2ma}$.
The asymptotic expression (\ref{FCnb1}) shows that the decay of the
topological FC for large proper lengths of compact dimensions is of power
law for both massless and massive fields. This behavior is in clear contrast
to the case of Minkowski bulk with the exponential decay for massive fields
(see (\ref{FCM})).

\begin{figure}[tbph]
\begin{centering}
\epsfig{figure=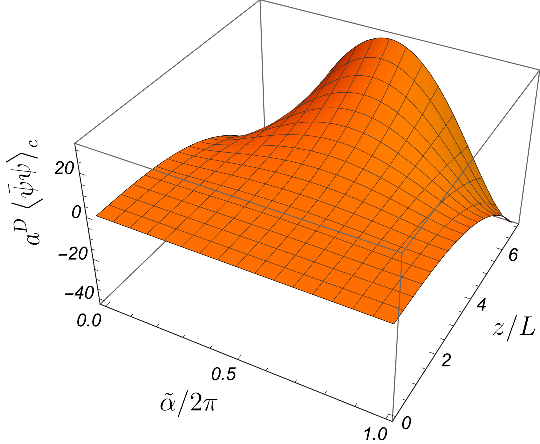,width=8cm,height=7cm}
\par\end{centering}
\caption{The topological FC in the model $(p,q)=(2,1)$ as a function of the
phase $\tilde{\protect\alpha}$ and the ratio $z/L$. The graph is plotted for 
$ma=1$.}
\label{fig1}
\end{figure}

The numerical analysis below will be given for the model with a single
compact dimension $x^{l}$, $(p,q)=(D-2,1)$ with the length $L_{l}=L$ and
with the phase $\tilde{\alpha}_{l}=\tilde{\alpha}$. The general formula (\ref%
{FCc}) is specified to 
\begin{equation}
\langle \bar{\psi}\psi \rangle _{\text{\textrm{c}}}=-\frac{Na^{-D}}{\left(
2\pi \right) ^{\frac{D+1}{2}}}\sum_{n=1}^{\infty }\cos \left( \tilde{\alpha}%
n\right) \,U_{ma}^{\frac{D+1}{2}}\left( 1+\frac{n^{2}L^{2}}{2z^{2}}\right) .
\label{FCcq1}
\end{equation}%
Figure \ref{fig1} displays the topological contribution of the FC (in units
of $a^{-D}$) in 4-dimensional space ($D=4$) for fixed $ma=1$, versus the
phase $\tilde{\alpha}$ and $z/L$. As it has been already clarified by
asymptotic analysis, the condensate tends to zero on the AdS boundary like $%
z^{D+1+2ma}$ and behaves as $z^{D-1}$ for large values of $z$ (region near
the horizon). The graph shows that the FC can be both positive and negative
depending on the phase $\tilde{\alpha}$.

\begin{figure}[tbph]
\begin{centering}
\epsfig{figure=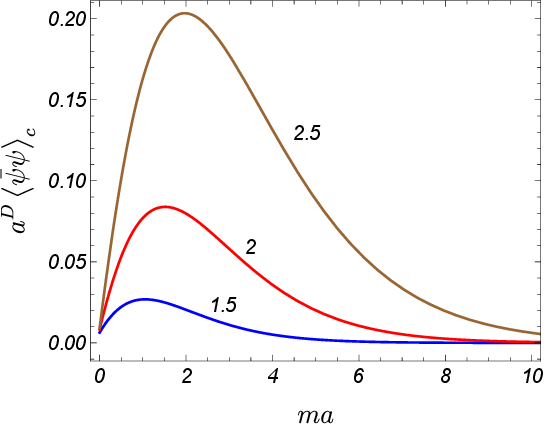,width=8cm,height=7cm}
\par\end{centering}
\caption{The topological FC versus the mass (in units $1/a$) for $\widetilde{%
\protect\alpha }=\protect\pi /2$. The numbers near the curves are the values
of $z/L$. }
\label{fig2}
\end{figure}

It is of interest to consider the dependence of the FC on the mass. That
dependence is presented in fig. \ref{fig2} for the model $(p,q)=(2,1)$ with $%
\tilde{\alpha}=\pi /2$. The graphs are plotted for $z/L=1.5;2;2.5$ (the
numbers near the curves) and show that the condensate is maximal for some
intermediate value of the mass. For large masses the topological FC is
exponentially suppressed as a function of $ma$.

\begin{figure}[tbph]
\begin{centering}
\epsfig{figure=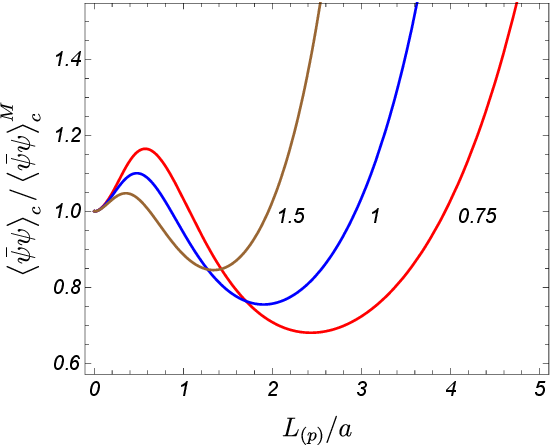,width=8cm,height=7cm}
\par\end{centering}
\caption{The ratio of the FCs for locally AdS and Minkowskian geometries as
a function of the proper length of the compact dimension (measured in units
of $a$) for $\widetilde{\protect\alpha }=\protect\pi /2$. The numbers near
the curves present the corresponding values of $ma$. }
\label{fig3}
\end{figure}

As it already has been discussed above, for a massive field in locally AdS
spacetime the dependence of the FC on the length of the compact dimension is
essentially different from that for locally Minkowski bulk. To compare those
dependences for the same values of the proper lengths of compact dimension
(in the same units of $a$), in fig. \ref{fig3} we have displayed the ratio $%
\langle \bar{\psi}\psi \rangle _{\text{\textrm{c}}}/\langle \bar{\psi}\psi
\rangle _{\text{\textrm{c}}}^{\text{\textrm{M}}}$ in the model $(p,q)=(2,1)$
as a function of $L_{\mathrm{(p)}}/a$, where $L_{\mathrm{(p)}}=L$ for the
Minkowskian case and $L_{\mathrm{(p)}}=aL/z$ for locally AdS bulk. The
graphs are plotted for $\tilde{\alpha}=\pi /2$ and $ma=0.75;1;1.5$ (numbers
near the curves). As seen, for large values of the proper length of compact
dimension $\langle \bar{\psi}\psi \rangle _{\text{\textrm{c}}}/\langle \bar{%
\psi}\psi \rangle _{\text{\textrm{c}}}^{\text{\textrm{M}}}\gg 1$. This is a
consequence of the power law decay of the AdS FC instead of an exponential
fall-off in the case of locally Minkowskian geometry. For small values of
the proper length compared with the curvature radius the effect of the
gravitational field is weak and the AdS FC tends to the one for locally
Minkowski bulk.

For even $D$ the Clifford algebra has two inequivalent irreducible
representations with the gamma matrices $\gamma _{\left( s\right) }^{\mu }$,
where $s=\pm 1$ enumerates those representations. We take $\gamma _{\left(
s\right) }^{\mu }=\gamma ^{\mu }$, $\mu =0,1,\ldots ,D-1$, with $\gamma
^{\mu }$ from (\ref{dmf}). For $\gamma _{(s)}^{D}$ one can take $\gamma
_{\left( s\right) }^{D}=s\gamma $, with $\gamma \equiv (a/z)^{D-1}\prod_{\mu
=0}^{D-1}\gamma ^{\mu }$, in the case $D=4n$ and $\gamma _{\left( s\right)
}^{D}=si\gamma $ in the case $D=4n+2$, where $n=0,1,2,\ldots $. Let $\psi
_{(s)}$ be the Dirac field with the Lagrangian density $\mathcal{L}_{(s)}=%
\bar{\psi}_{\left( s\right) }[i\gamma _{\left( s\right) }^{\mu }(\partial
_{\mu }+\Gamma _{\mu }^{\left( s\right) })-m]\psi _{\left( s\right) }$,
realizing the representation for a given $s$. The mass term breaks the
parity ($P$-) and charge ($C$-) invariances for $D=4n$ and the time reversal
($T$-) invariance for $D=4n+2$. The fermionic models with $C$-, $P$- and $T$%
-symmetries are constructed combining the separate fields with appropriate
transformation rules. Introducing new fields $\psi _{\left( s\right)
}^{\prime }$ in accordance with $\psi _{\left( +1\right) }^{\prime }=\psi
_{\left( +1\right) }$ and $\psi _{\left( -1\right) }^{\prime }=(a/z)\gamma
\psi _{\left( -1\right) }$, the combined Lagrangian density $\mathcal{L}%
=\sum_{s}\mathcal{L}_{(s)}$ is written in the form $\mathcal{L}=\sum_{s}\bar{%
\psi}_{\left( s\right) }^{\prime }[i\gamma ^{\mu }(\partial _{\mu }+\Gamma
_{\mu })-sm]\psi _{\left( s\right) }^{\prime }$, with $\gamma ^{\mu }=\gamma
_{(+1)}^{\mu }$ being the gamma matrices we have used in the discussion
above. From here it follows that the topological contributions in the FC for
the fields $\psi _{\left( s\right) }^{\prime }$ are given by $\langle \bar{%
\psi}_{\left( s\right) }^{\prime }\psi _{\left( s\right) }^{\prime }\rangle
_{\text{\textrm{c}}}=s\langle \bar{\psi}\psi \rangle _{\text{\textrm{c}}}$,
where $\langle \bar{\psi}\psi \rangle _{\text{\textrm{c}}}$ is the FC
investigated above for the case $s=+1$. Now we can return to the initial
fields $\psi _{(s)}$. By using the relation $\gamma ^{(0)}\gamma ^{\dag
}\gamma ^{(0)}\gamma =-(z/a)^{2}$, for the corresponding FCs one obtains $%
\langle \bar{\psi}_{\left( s\right) }\psi _{\left( s\right) }\rangle _{\text{%
\textrm{c}}}=s\langle \bar{\psi}_{\left( s\right) }^{\prime }\psi _{\left(
s\right) }^{\prime }\rangle _{\text{\textrm{c}}}$. Hence, we get $\langle 
\bar{\psi}_{\left( s\right) }\psi _{\left( s\right) }\rangle _{\text{\textrm{%
c}}}=\langle \bar{\psi}\psi \rangle _{\text{\textrm{c}}}$ and the
topological FCs coincide for the fields realizing two inequivalent
representations of the Clifford algebra and they are given by general
formula (\ref{FCc}).

An interesting area of {}{}applications of fermionic models is 2D Dirac
materials \cite{Wehl14}. These are planar condensed matter systems in which
the long-wavelength excitations of the electronic subsystem are described by
the Dirac equation in two-dimensional space. An example of such materials is
graphene. In graphene, the components of the spinors $\psi _{(s)}$
correspond to the amplitude of $\pi $-electron wave function on the
triangular sublattices and $s=\pm 1$ correspond to the points $\mathbf{K}%
_{\pm }$ of the Brillouin zone. For graphene nanotubes the background
spacetime of the effective Dirac model is flat with spatial topology $%
R^{1}\times S^{1}$. This is the flat spacetime analog of the model we have
considered above for $D=2$ and $(p,q)=(0,1)$. In graphene nanotubes the
phase $\alpha _{1}$ in the periodicity condition (\ref{qpc}) depends on the
chirality of the tube. In metallic nanotubes $\alpha _{1}=0$ and in
semiconducting nanotubes $\alpha _{1}=\pm 2\pi /3$ for $s=\pm 1$. In the
absence of magnetic flux threading the semiconducting tube the FCs from the
spinors $s=\pm 1$ coincide. In the presence of magnetic flux one has $\tilde{%
\alpha}_{1}=\alpha _{1}-2\pi \Phi _{1}/\Phi _{0}$ and the FC is different
for negative and positive values of the phase $\alpha _{1}$. An effective
background curvature in Dirac models describing the condensed matter systems
can be generated, for example, by local variations of the Fermi velocity of
electrons, by external fields, and by deformations of the lattice (see,
e.g., \cite{Amor16}).

\section{Conclusion}

\label{sec:Conc}

We have studied the combined effects of background curvature and nontrivial
topology on the FC in locally AdS spacetime with toroidally compactified
spatial dimensions in Poincar\'{e} coordinates. The Casimir contribution $%
\langle \bar{\psi}\psi \rangle _{\text{\textrm{c}}}$ is explicitly extracted
and the renormalization of the total FC is reduced to the one in AdS
spacetime. The topological part in the FC is given by (\ref{FCc}) with the
function $U_{\nu }^{\mu }\left( x\right) $ defined in (\ref{pQ}). A simpler
representation for the latter is obtained for odd number of spatial
dimensions (see (\ref{pu2})). For a given spatial dimension, the combination 
$a^{D}\langle \bar{\psi}\psi \rangle _{\mathrm{c}}$ (FC measured in units of 
$a^{-D}$) is completely determined by the phases $\tilde{\alpha}_{l}$ and
the ratios $L_{\mathrm{(p)}l}/a$ being the proper lengths of compact
dimensions in units of the curvature radius. The effect of the phases $%
\alpha _{l}$ in the periodicity conditions (\ref{qpc}) can be interpreted in
terms of magnetic fluxes enclosed by compact dimensions and vice versa. The
topological FC is an even function of the phases $\tilde{\alpha}_{l}$ and
may be either positive or negative, depending on their values.

For small proper lengths of compact dimensions, compared to the spacetime
curvature radius, the influence of the gravitational field on the
topological FC is weak and the leading term in the corresponding expansion
coincides with the FC in locally Minkowski spacetime. For given values of
the coordinate lengths $L_{l}$, this corresponds to the region near the AdS
horizon with the behavior $\langle \bar{\psi}\psi \rangle _{\text{\textrm{c}}%
}\propto z^{D-1}$. The effect of gravity is crucial for proper lengths of
the order or larger than the curvature radius. In particular, for $L_{%
\mathrm{(p)}l}/a\gg 1$ the decay of $\langle \bar{\psi}\psi \rangle _{\text{%
\textrm{c}}}$, as a function of $L_{\mathrm{(p)}l}$, follows a power law,
like $1/L_{\mathrm{(p)}l}^{D+1+2ma}$. In the problem on locally Minkowski
bulk the corresponding suppression of the FC for massive fields is
exponential. For given $L_{l}$, the Casimir contribution to the FC tends to
zero on the AdS boundary as $z^{D+1+2ma}$. The purely AdS part $\langle \bar{%
\psi}\psi \rangle _{0}$ does not depend on $z$ and it dominates in the
region near the boundary.

For even values of $D$ the discussion presented above covers both the
irreducible representations of the Clifford algebra. The FCs for the fields
realizing those representations coincide if their masses and periodicity
conditions are the same. The 2D materials with long-wavelength properties
described by the Dirac model are among condensed matter applications of the
results presented in the paper.

\section*{Acknowledgments}

A.A.S. and T.A.P. were supported by the Grants No. 21AG-1C047 and No.
24FP-3B021 of the Higher Education and Science Committee of the Ministry of
Education, Science, Culture and Sport RA.

\end{document}